\documentstyle[sprocl]{article}

\input{psfig}
\input{epsfig.sty}

\def\be{\begin{equation}}
\def\ee{\end{equation}}
\def\bea{\begin{eqnarray}}
\def\eea{\end{eqnarray}}

\begin{document}

\title{Sensitive Search for a Permanent Muon Electric Dipole
Moment.\footnote{Submitted for  
publication in Proceedings of the International Workshop on High
Intensity Muon Sources (HIMUS99), KEK, Japan, December 1-4 1999.}}

\author{Y.K. Semertzidis, H. Brown, G.T. Danby, J.W. Jackson, R. Larsen,\\
D.M. Lazarus, W. Meng, W.M. Morse,  C. Ozben, R. Prigl}

\address{Brookhaven National Lab, Upton, New York 11973} 

\author{R.M. Carey, J.P. Miller, O. Rind, B.L. Roberts, L.R. Sulak}

\address{Department of Physics, Boston University, Boston,
Massachusetts 02215}

\author{ V. Balakin, A. Bazhan, A. Dudnikov, B.I. Khazin, G. Sylvestrov}

\address{Budker Institute of Nuclear Physics, Novosibirsk, Russia}

\author{Y. Orlov}

\address{Newman Laboratory, Cornell University, Ithaca, New York 14853}

\author{K. Jungmann}

\address{Physikalisches Institut der Universitat Heidelberg, 69120
Heidelberg, Germany}

\author{P.T. Debevec, D.W. Hertzog, C.J.G. Onderwater}

\address{Physics Department, University of Illinois at
Urbana-Champaign, Urbana-Champaign, Illinois 61801}

\author{E.J. Stephenson}

\address{Indiana University Cyclotron Facility, Bloomington, Indiana 47408}

\author{P. Cushman, I. Kronkvist}

\address{Department of Physics, University of Minnesota, Minneapolis,
Minnesota 55455}

\author{F.J.M. Farley}

\address{Department of Physics, Yale University, New Haven,
Connecticut 06511}


\maketitle\abstracts{ We are proposing a new method
 to carry out  a dedicated search for a 
permanent electric dipole moment (EDM) of the muon with a sensitivity
at a level of $10^{-24} \, \rm e \cdot cm$. The experimental  design
exploits the strong 
motional electric field sensed by relativistic particles in a magnetic 
storage ring \cite{muon_loi_97,Yannis_98}. As a key feature, a novel technique 
has been invented in which the g-2 precession is compensated with 
radial electric field.  This technique will benefit greatly when the
intense muon sources advocated by the developers of the muon storage
rings and the muon colliders become available.}

\section{Motivation}

The standard model of particle physics is a very successful
theoretical framework, 
which describes all confirmed observations to date. 
 However, the model leaves important questions concerning the
physical nature of observed processes unexplained, although it provides
an accurate description of them. Among the not yet understood 
phenomena are the reasons for parity violation, 
the particle masses, the violation of the CP symmetry. 
CP violation  is the only known mechanism that could explain the
matter antimatter asymmetry found in the universe.
In order to obtain a deeper insight, speculative models have been
suggested which often are 
connected to observable deviations from standard theory predictions,
particularly 
violations of assumed symmetries or yet unknown properties of particles.
The spectrum of these theories includes
super symmetry \cite{Bern_91,Chang_99}, left-right symmetry~\cite{Gen_90},
multi Higgs scenarios \cite{Bar_97,Bow_97} and many other important approaches.
This strongly motivates sensitive searches for
forbidden decays, e.g. the lepton number violating muon electron conversion 
\cite{MECO_98} , the decay  $\mu \rightarrow e \gamma$ \cite{MuEG_99} 
and precise measurements of particle characteristics such as the muon
magnetic moment anomaly 
\cite{Carey_99}.  

A fundamental particle is described  by only a few parameters such as its mass,
its intrinsic angular momentum, its electric charge (electric
monopole moment),  
its magnetic dipole moment, and a set of  conserved quantum numbers,
like lepton flavor 
\cite{PDG}.
A permanent EDM has not been observed for any of them.
It would
violate both parity (P) and time reversal (T) invariance. 
If CPT is assumed to be a valid unbroken symmetry, a permanent
EDM would hence be a signature of CP violation \cite{Barr_89,Bern_91}.  

The standard model of particle physics predicts 
a CP violating EDM in fundamental particles at the multi loop level of
amplitude more than 
five orders of 
magnitude below the sensitivity of present experiments \cite{Pos_94} 
(see Table 1).
Therefore searches for a permanent particle EDM render excellent
opportunities to 
test models beyond standard theory where in some cases they predict effects 
as large as the presently known experimental bounds. 
Despite the non-observation of a positive EDM
signal such research has nevertheless been most successful in steering
the development of theoretical particle physics over many decades
and also was the incentive for the ever increasing precision
in the experiments themselves.
The absence of any observed finite EDM for the neutron, for example, 
has disfavored more speculative models than any other experimental
approach so far \cite{Ram_99}.                                        

Searches for EDMs have been performed with highest sensitivity  
for electrons~\cite{Commins_94} (e) and neutrons~\cite{Harris_99}
(n). Further there are experimental limits for   
muons~\cite{Bailey_78} ($\mu$), tauons~\cite{Aciari_98} ($\tau$) and
protons~\cite{Cho_89} ($p$)  
(Table 1). 
The experiments
include beams of neutral atoms, molecular beams, stored neutrons and 
stored charged particles. 
In heavy atoms and particularly in polar molecules there are substantial 
enhancement factors due to the utilization of the rather strong internal 
electric fields within these systems \cite{Sandars_67}.

\begin{table}[t]
\caption{Limits on Electric Dipole Moments $d$ for 
         electrons ($e$),
         muons ($\mu$),
         tauons ($\tau$),
         protons ($p$) and 
         neutrons ($n$).
         For various models the scaling with the lepton mass 
        is stronger than linear (x$>$0).  The number in the first
parenthesis for the electron case refers to the statistical error and
the second to the systematic.}
\vspace{0.2cm}
\begin{center}
\footnotesize
\begin{tabular}{|c|rl|c|rl|} \hline
      & Present Limit & on  $|d|$            & Standard Model
    & New&Physics           \\ 
        &[$10^{-27} \, \rm e \cdot cm$] &  &  Prediction [$10^{-27} \,
    \rm e \cdot cm$] & Limits & [$10^{-27} \, \rm e \cdot cm$] \\
    \hline\hline 
$e$        & 1.8\,(1.2)\,(1.0) &                     &
    $\stackrel{\large <}{\small\sim}10^{-11}$  
                                                   &$\stackrel{\large
    <}{\small\sim}1$&\\ 
\hline 
$\mu$    & $<1.05\cdot10^9$&\small{(95\% C.L.)}   & $\stackrel{\large
    <}{\small\sim}10^{-8}$  
                                                   &$\stackrel{\large
<}{\small\sim}200$&$\times(\frac{m_{\mu}}{m_e})^x$\\ \hline                   
                                                                            
$\tau$   & $<3.1\cdot10^{11}$&\small{(95\% C.L.)} & $\stackrel{\large
<}{\small\sim}10^{-7}$                                                    
&$\stackrel{\large <}{\small\sim}1700$&$\times(\frac{m_{\tau}}{m_e})^x$\\
\hline  p        & $-3.7\,(6.3)\cdot10^4$&                & $\sim10^{-4}$  &
$\stackrel{\large <}{\small\sim}60$&\\ \hline n        &
$<63\cdot10^0$&\small{(90\% C.L.)}              & $\sim10^{-4}$  &
$\stackrel{\large <}{\small\sim}60$&\\ \hline
\end{tabular}
\end{center}
\end{table}

The upper bound extracted from electron and neutron experiments
already disfavor super-symmetric models  with CP violating phases of
order unity and suggest variants with phases of order $\alpha/\pi$
\cite{Dim_95,Bar_96}. 
Other significant restrictions of the parameter space would also be an
option, however 
at a loss of generality.
A muon EDM experiment at $10^{-24} \, \rm e \cdot cm $ will be
competitive here. 
Moreover, it will provide valuable complementary information,
because the muon belongs to the second generation of particles
where in the quark sector CP violation occurs. 
In two Higgs doublet models \cite{Bar_97} with large 
ratios for the vacuum expectation values of the involved Higgs fields
($\tan\beta$ ($\leq 15$))  
and for left right symmetry a muon EDM could be as large as a few
$10^{-24} \, \rm e \cdot cm $
\cite{Gen_90} and some special models allow values up to $10^{-21}\,
\rm e \cdot cm $~\cite{Gen_90}.

Sensitive searches are presently being carried out
on neutral objects, i.e. neutrons, atoms and molecules. 
This choice was strongly influenced by the Ramsey-Purcell-Schiff 
theorem \cite{Schiff_63} which states that for point-like charged objects
in electromagnetic equilibrium, the net electric field averages to 
zero. The widely known loopholes  so far were weak and strong nuclear 
forces, weak electron-nucleon forces and relativistic forces.
It is recognized now that this theorem is also 
not applicable to particles in a storage ring, particularly to
the method proposed here, where 
motional fields are employed, because it is not possible
to factorize particle velocity and electric field, which constituted the
basis of the theorem \cite{Sandars_99,Schiff_63}.
Therefore these electric fields, which are very strong for 
relativistic particles, can be beneficially exploited. Such fields 
can be three orders of magnitude larger than technically achievable 
fields between electrodes, where 5.5~MV/m are regarded an upper limit.

      It should be mentioned that the muon is the only second generation
      particle for which a precision EDM experiment is
feasible~\cite{pilaftsis}. Since CP 
      violation is associated with the second and third generation of
quarks only, 
      the muon may have a unique window for new physics in the lepton sector.
      Therefore, even if there were an EDM observed in another system,
a measurement  
      on the muon would be extremely important to understand the
nature of the effect 
      \cite{Babu_98}.                                                        
      
      The Muon g-2 Experiment, E821, now being conducted at BNL, has
been designed to probe physics 
beyond the standard model~\cite{E821}. 
      This includes super-symmetry with large $\tan \beta$ 
      \cite{Nath_99,Chat_96,Mor_96,Her_99,Kin_90}
      where the muon EDM also has sensitivity. The experiments
complement each other, 
      because the magnetic anomaly and the EDM are related to each other
      as real and imaginary parts of the same physical
      quantity \cite{Mar_99}. The recent new limit on the muon
magnetic anomaly  
      \cite{BLR_00}
      corresponds (scaling by $e/m_\mu$ and assuming the natural CP-violating 
phase to be of order 1)
      to a dipole moment of order $5 \times 10^{-22} \, \rm e \cdot
cm$~\cite{Mar_99a}, which is  
      well within the reach of this experiment, and demonstrates, how  
      the explored parameter space can be expanded.

\section{Experimental Overview}

The muon spin precession angular frequency (relative to the momentum vector) 
in the presence of both electric and magnetic fields is given by:

\begin{equation}
\vec{\omega}  = -{e \over m} \{a \vec{B}  + ({1 \over{\gamma^2 -1}} - a)
{{\vec{\beta} \times \vec{E} } \over c} + 
 {\eta \over 2} ({\vec{E} \over c} + \vec{\beta} \times \vec{B}) \} , \label{eq:spin2}
\end{equation}

\noindent
(assuming the 
muon velocity is orthogonal to the external magnetic and electric fields $\vec{\beta}
\cdot \vec{B} =\vec{\beta}
\cdot \vec{E} = 0$)
where $a=(g-2)/2$ and $\eta$ is the EDM in units of 
${e\hbar}\over {4mc}$.  $\eta$ plays a role for the EDM corresponding to  
the g factor for the magnetic dipole moment. The muon EDM couples to the
external magnetic field because in the muon rest frame it looks like an
electric field due to the Lorentz transformation.  The EDM value in MKS units 
is given by 

\begin{equation}
d_{\mu} = {\eta  \over 2} {e \hbar \over 2 m c} \simeq \eta \times 4.7 \times
10^{-14} \, {\rm e \cdot cm}.         \label{eq:edmmu}
\end{equation}

From the above equation~(\ref{eq:spin2}) is apparent why the present $g-2$ 
experiment uses muons at the ``magic'' $\gamma$, i.e. $\gamma = 29.3$ since
at this value the coefficient ${1 \over{\gamma^2 -1}} - a = 0$.
and the muon spin precession becomes independent of the electric
field present in the lab frame.  For the dedicated EDM experiment we propose 
to use muons with much lower energy as well as a radial electric field which
cancels the g-2 precession leaving the EDM precession to operate on its own
and accumulate over many microseconds.  It should be noted that the
main component which 
causes the spin to precess vertically in the muon rest frame is the
dipole magnetic field which is partially transformed into a radial electric
field in the muon rest frame.  The electric field in the lab required
to cancel the g-2 
precession is

\begin{equation}
E = {{a B c (\gamma^2 - 1) } \over {\beta}} = 2 \,{\rm MV/m} ,
 \label{eq:efield}
\end{equation}

\noindent{for} $\gamma=5$, $B=0.24\,{\rm T}$ and a magnet radius of $711 \, 
{\rm cm}$.  
Concentric cylindrical plates $10 \, {\rm cm}$ apart with $\pm 100$~KV
voltage will satisfy this condition.

From  equation~(\ref{eq:spin2}) (at the muon ``magic'' momentum) it is 
apparent that the EDM influence on the
spin precession is two-fold: 1) The spin precesses about
an axis which is not exactly parallel to the magnetic field but is at an
angle $\delta = \tan^{-1}{\eta \beta \over 2 a}$.  The maximum excursion 
of the 
muon spin from the horizontal plane, due to an EDM, occurs when the spin is at 
$90^{\circ}$ with respect to the momentum vector.  When the spin and momentum
vectors are aligned there is no excursion from the horizontal plane.  
2) The spin precession frequency is increased by a factor of
 $\sqrt{(1 + \delta^2)}$.

Both of the above methods were used in the last CERN muon g-2 experiment to set
a limit on the EDM~\cite{Bailey_78}.  They searched for an up down
asymmetry in  
the counting rate with a detector split on the vertical mid-plane.  
The presence of an EDM induces 
a difference between the counting rate of the two parts of the split detector
at the anomalous spin precession frequency.  
The reported limit is $3.7 \pm 3.4 \times 10^{-19} \rm \, e \cdot cm $.
The error is the combined result of the statistical and
systematic errors:

\begin{itemize}

\item Statistical: $\pm 2.7 \times 10^{-19} \rm \, e \cdot cm $

\item Systematic: $ \pm 2 \times 10^{-19} \rm \, e \cdot cm $

\end{itemize}

\noindent
Both errors are at the same level.  There are at least two sources of
systematic errors when the spin is allowed to precess:  1) The center
of the muon population is not exactly aligned with the geometric
center of the detectors which 
results to an up-down modulation of the average decay positron
position with the g-2 frequency.   2) As the muon spin precesses, the
decay positrons travel 
through different magnetic and electric fields before they reach the
detector which results in an up-down modulation of the average
decay positron 
position with the g-2 frequency.

\begin{figure}[t]
\label{Fig1}
 %
 \begin{minipage}{1.75in}
  \centering{
   \hspace*{-0.3in}
   \mbox{
   \epsfig{figure=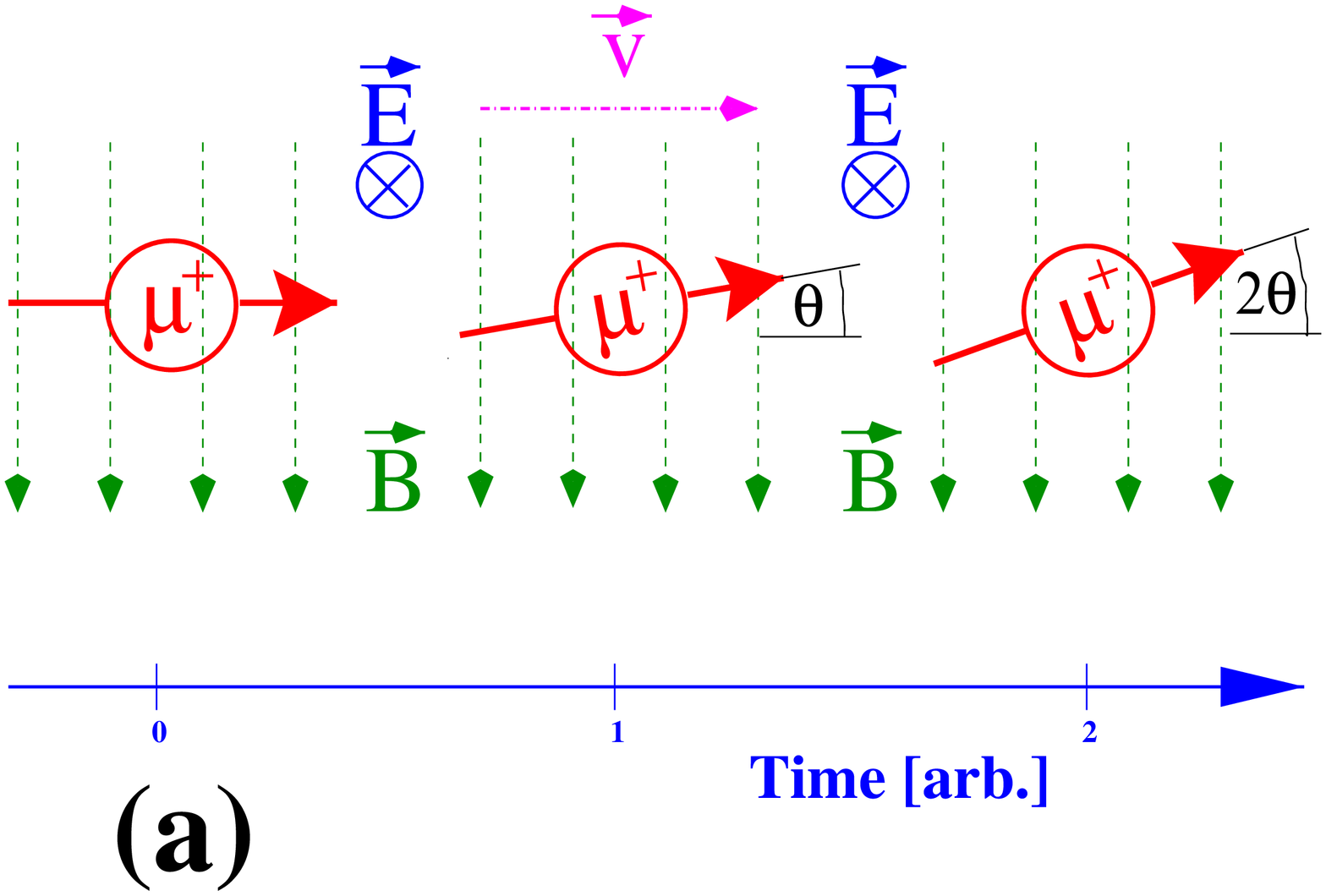,width=2.0in,angle=270}
         }
             }
 \end{minipage}
 \hspace*{0.1cm}
 \begin{minipage}{1.75in}
  \centering{
   \hspace*{-0.0in}
   \mbox{
   \epsfig{figure=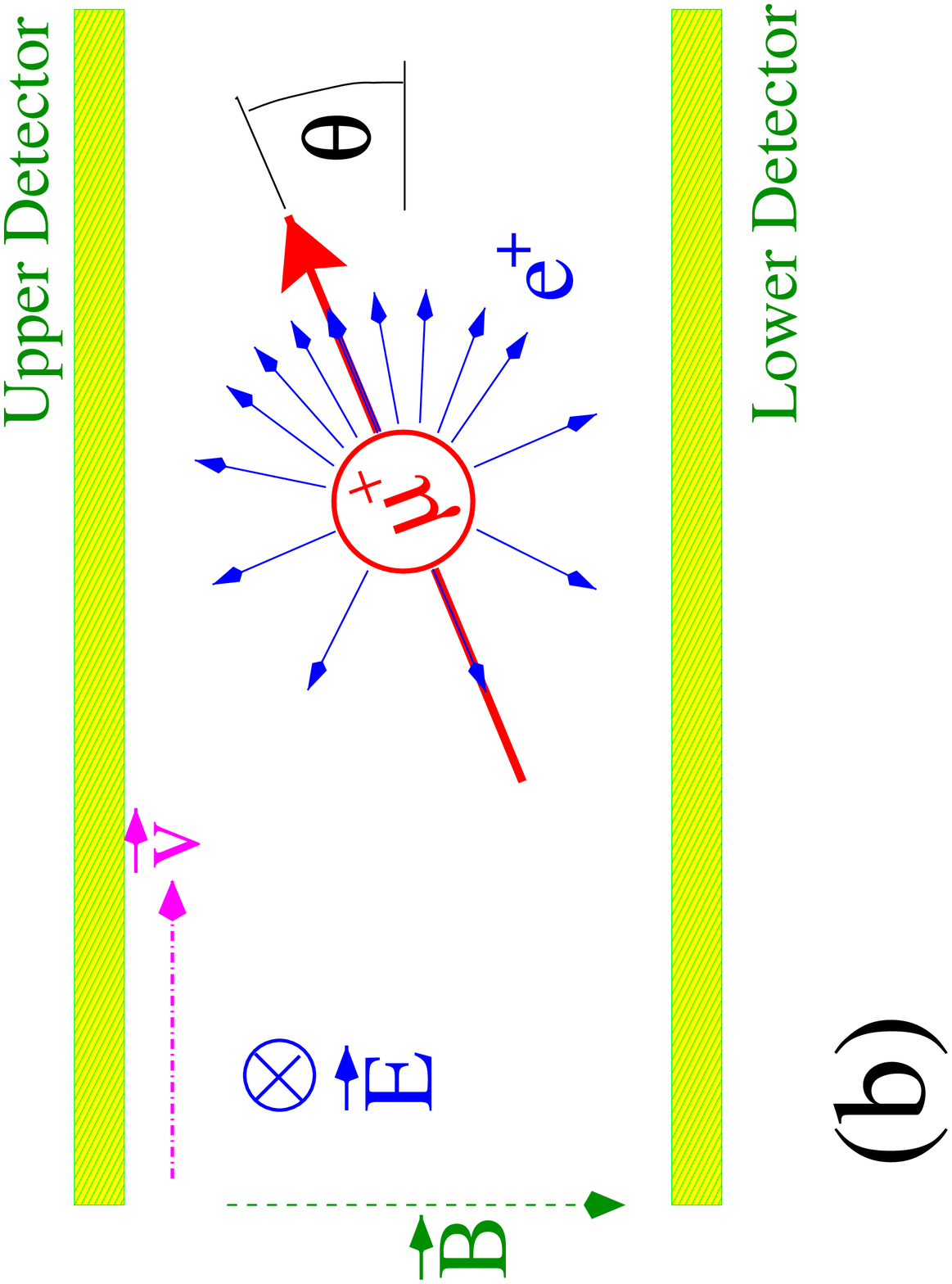,width=2.0in,angle=270}
         }
             }
   \end{minipage}
 \centering\caption[]
        {{The basic principle of the proposed experiment.
         (a) Relativistic muons ($\mu^+$) of velocity $\vec{v}$
         moving in a magnetic storage ring with field ($\vec{B}$) find
themselves   
         exposed to a motional electric field $\vec{E} \propto \vec{v}
\times \vec{B}$. 
         In case of a finite small EDM the muon spin precesses with a linear 
         increase in precession angle
         $\theta$ in time about an axis which is directed
radially. (b) Due to the spatial anisotropy in the 
decay $\mu \rightarrow e^+ 
         + \nu_e + \overline{\nu_{\mu}}$ detectors above and below the
storage region are expected 
         to observe a time dependent change in the ratio of positron
counting signals. The  
         positron angular distribution is indicated by the density of
arrows.    
       \label{fg:principle}} }

\end{figure}

We propose to use a radial electric field to cancel the 
$g-2$ precession.  This together with other improvements will significantly
reduce many systematic errors and improve the experimental sensitivity to 
a muon EDM.  The muon spin direction will be ``frozen'' horizontally.
In the presence of a non-zero EDM, 
the proposed radial E-field (i.e. the motional E-field in the muon
rest frame due to the laboratory dipole B-field) will rotate the spin around
an axis parallel to the E-field, thereby tipping the spin vertically.
This results in an up-down asymmetry in the number of electrons
which is growing linearly with time and
this is the quantity which we will measure (see Fig.~\ref{fg:principle} ).

Other changes from the present $g-2$ experiment include:

\begin{enumerate}

\item{} Use {\it magnetic} focusing instead of electrostatic focusing as is
used in the present g-2 ring.  A 10 by 10~cm area is available for
muon storage which will increase the number of stored muons substantially.

\item{} Eliminate on the average all, out of plane (``vertical''), electric
fields from the system during normal  
data taking.  Any radial magnetic field in the presence of compensating
vertical electric
field introduces a systematic limitation~\cite{orlov,jim}.  For example, in 
the g-2 ring, a net radial magnetic field of 10~ppm of the main field
would produce 
a systematic error in the EDM of $1 \times 10^{-21}\,{\rm e \cdot cm}$.  A 
vertical electric field can be used, as discussed later in this note,
to calibrate the system sensitivity with high accuracy.

\end{enumerate}

\subsection{Storage Ring-Weak Magnetic Focusing}

The g-2 ring magnet has been built to provide excellent field stability
and homogeneity for precision measurements. For the EDM experiment
the field homogeneity is of relatively little concern, but the stability
of the magnetic field, specifically its direction, is relevant.  It is
relevant, however, only
as a second order effect depending on the E-field homogeneity as a function of 
the average vertical position of the muon beam.
Measurements of the minor field components were done with Hall probes to
an accuracy of about 10 ppm. During data taking runs the maximum
number of muons is stored if the beam is vertically centered with
respect to the electrostatic quadrupoles. Current coils on the pole faces
are used to compensate the static radial field average. From these
measurements we know that the average radial field $<B_r>$ changes by
$20 \pm 10$~ppm year to year which is more than adequate for an
 E-field directional homogeneity of $10 \, \rm \mu rad/10cm$.
The E-field homogeneity can be easily studied by raising and lowering
the beam by inducing a radial B-field and looking at the vertical muon 
spin precession as a function of the vertical beam position.

From the NMR measurements of the absolute field, we
know that the variations in the magnetic field are dominated by changes
in the vertical gap between the upper and lower pole pieces due to
thermal expansion/contraction of the magnet steel including the super-bolts
that hold the yoke pieces together. A uniform temperature rise leads
to an increase of the air gap and thus a reduction of the magnetic field
in the storage region but does not change the direction of the field.
However, the thermal time constant of the yoke is several hours, and
changes in the ambient temperature do not only change the average vertical
gap but also cause a transient relative tilt of the top pole with respect
to the bottom. This affects the direction of the magnetic field,
introducing a small radial field component, as well as
the normal quadrupole
moment, i.e. the gradient of the vertical field along the radial direction.
In a weak focusing magnet
we can use NMR probes to keep track of the radial field change by
precisely measuring the normal quadrupole moment. These measurements
should be compared to direct measurements of the gap variation using
optical interferometry. During the first g-2 run in 1997 data were taken
in a constant current mode and the ambient temperature changes are
clearly visible in the average field. This change is highly correlated
with the 
temperature difference between the magnet yoke surface temperature and
the ambient temperature.
Since 1997 the magnet yoke has been insulated and the temperature effects
are reduced by about an order of magnitude,
but the radial field component
could still change by a few hundred nanoradians over a day which is
tolerable.  With
further improvements in the temperature stability, this change will
be even smaller.

Most of the pulsed NMR equipment for g-2 could be used in the EDM
experiment. Only the passive probes and the pulse amplifiers have a narrow
bandwidth and would have to be replaced.

 For the
B-field there is a strong dependence due to the attractive
magnetic force. At the g-2 field of 1.45 Tesla the average gap is
reduced by 0.3 mm relative to the zero field value and the
top pole rotates by about 0.3 mrad with respect to the bottom
pole during the ramp. This translates into a tilt change of about
1.0 $\mu rad$ or a change in the radial field component in the
mid-plane by 0.5 ppm while going from (e.g.) 0.218 T  to 0.233 T .
Again this change can be monitored accurately with NMR probes and
also with optical interferometers. Using both methods would give
additional information on field perturbations due to permeability
changes which are expected to be small at these low field values.

There is a need to have a kicker in order to be able to inject muons into the
ring.  One third in length of the current kicker with two thirds the peak 
current would
suffice.  The eddy currents of this kicker would be sufficiently low not
to become a source of systematic errors.

\subsection{Detectors}

We require a detector system which is sensitive to the number
of muon decay electrons going upward compared to the number going downward,
in an energy range of 100 to 500 MeV.
The up-down asymmetry, ${N_{up}-N_{dn} \over N_{up}+N_{dn}}$,
is directly proportional to $\eta$ and the EDM.
The detector should be relatively insensitive to background from muons,
neutrons, pions, protons, etc.

The detectors will also have to be capable of handling the extremely
high instantaneous rates immediately following muon injection,
and the efficiency for measuring the up-down asymmetry
must remain stable when the rate diminishes as the muons decay.

We anticipate that $10^7$ muons will be stored per AGS bunch, significantly
more than, for example, what the muon g-2 experiment currently receives.
There will be 12 bunches, separated by about $33 ms$, as in the g-2
experiment.

The proposed system consists of 2~cm thick scintillating lead-glass
slabs placed above 
and below and on the sides of the storage region, with photodiode readout.
The radiation length is about 1.5 cm. The electrons, on average, enter the
lead-glass at large angles with respect to the normal to the detector
surface.  Simulations show that the rms energy 
resolution for 400 MeV electrons is about 14\% due to shower fluctuations
and losses. Photon statistics will increase this; for
regular lead-glass, the resolution will increase to about 25\%, less if
scintillating lead-glass is used.
About 60\% of the electrons above 100 MeV fall into the main shower
peak, with the rest falling in a low-energy tail.
On average, the showering electrons will leave much more energy than,
say, muons which are lost from storage or gamma rays from neutron
capture, thereby minimizing background sensitivity.

The lead-glass slabs would cover the top and bottom areas of the pole
tips, and will go  around
the 45~m circumference of the ring minus 4~m for the required
gaps.
It will be segmented in 10~cm sections,
each section being read out by a photodiode. This provides an
overall segmentation of about 800.

The photodiodes will have time constants of about 1 $\mu s$. At the high
anticipated rates, individual counting of decay electrons
will not be attempted. Instead, we will integrate the
total charge accumulated in approximately 1 $\mu s$ intervals,
as a function of time
after injection. The charge will be calibrated to the number of decay
electrons, and the up-down ratio as a function
of time after injection will be deduced.

Individual electron counting is impractical at early times because the time
spacing between events is too small. Consequently, we plan
to integrate the total charge from the detectors. This implies that the
gains of the detectors must be known very well. The geometric
arrangement of the counters will be up-down symmetric so as to
minimize up-down asymmetries arising from backgrounds.

The measurements will be taken for about ten muon lifetimes after injection,
or for about 110 microseconds, however, most of the data will be
collected in the first 55 $\mu s$. 
The gains of the photodiodes must be extremely good over 55 $\mu s$:
to about 1 part in $10^5$ or known to that level. We have outlined two 
approaches to achieve this:  1) We are working with the
Instrumentation Division of BNL
at BNL to develop such an ultra-stable system. Their first analysis indicates
that one can stabilize the gains to 1 part in $10^4$ and then
calibrate the gain to 10\%, thus providing the needed 1 part in $10^5$
stability.  2)  We will calibrate the system in situ by observing the
muon spin precession. The resulting waveform should be an ideal
sine-wave multiplied by an exponential decay.  Any discrepancy from this 
ideal waveform can be assigned to gain changes and corrected for.

\section{Sensitivity Level}
The EDM sensitivity is given by (in MKS units)

\begin{equation}
\eta = {2 \theta_t  \over \beta B t }{m \over e},   \label{eq:eta}
\end{equation}

\noindent{with} $\theta_t$ the minimum vertical spin precession angle which
can be detected, $m, \,e$
the muon mass and charge respectively, and $B$, the average magnetic field seen
by the muon.  The uncertainty in $\eta$ is 

\begin{equation}
\sigma_{\eta} = {1  \over \gamma \tau_0 A_1 A \sqrt{2 N_{Tot} } },   
\label{eq:uneta}
\end{equation}

\noindent{with} $A_1 = \beta e B / 2 m = 10^8 / {\rm s}$ for 
$B = 0.24 {\rm T}$, $N_{Tot}$ the total 
number of observed decays, $\tau_0$ the muon lifetime at rest, $\gamma$ the 
Lorentz boost factor, and A the asymmetry.
If one observes
$1.5 \times 10^{15}$ decay electrons which have an average transverse 
asymmetry 
of 0.3 then
the uncertainty in  $\eta$
is $ 5.5\times10^{-11}$, and on the EDM  $2.5 \times 10^{-24} \, {\rm
e \cdot cm}$. 
For two muon lifetimes (i.e. $2 \times 11 {\rm \mu s}$) this corresponds to a 
vertical 
precession angle of 120~nrad.

The sensitivity of the experiment increases as the B field and the lifetime 
increase.  However, the necessary electric field required to cancel the
muon $g-2$ spin precession also grows fast as can be seen by 
equation~(\ref{eq:efield}).

\subsection{Calibration of Sensitivity}
A simple way to calibrate the experimental sensitivity to the muon
EDM is to apply a vertical electric field in the muon storage region.  The
beam will be displaced vertically until the magnetic focusing will compensate
with an equal and opposite force.  The average radial magnetic field seen by 
the muons will be such that it compensates the vertical electric field.  
This non-zero radial magnetic field will induce a non-zero spin precession
in the vertical plane mimicking an EDM effect.

The vertical spin precession angle with an out of plane (``vertical'')
 electric field $E_v$, 
and a vertical $B$ magnetic field present is given by

\begin{equation}
\theta_t = {e \over 2 m}[g {E_v \over c \beta \gamma^2} +
 \eta \beta B] t,           \label{eq:cal}
\end{equation}

\noindent{which} for an average $E_v = 5$~KV/m (averaged over the
circumference) results in 30~mrad vertical spin
precession which is easily 
measurable.  This average electric field can be achieved by placing  pulsed
25~KV on a pair of electrodes, 10~cm apart, covering $2\%$ of the ring. 

It should be noted that in the absence of an electric field the average 
radial magnetic field seen by the muons is exactly zero since we are using
 {\it magnetic} focusing.

\section{Systematic Errors}
Potential systematic errors include the following:

\begin{enumerate}

\item{}   Early to late electron counting (rate) effects.  With almost 
$10^7$ muons entering the ring and a good fraction of them decaying within 
$10 \, {\mu s}$ the rates are very high.  Our detector will be scintillating 
lead-glass calorimeters with photodiode readout.  The readout
electronics will be 
specially made so that they maintain their linearity to
$10^{-4}$ and will be measured~\cite{VR} to $10 \%$.
With 800 such systems the error is further reduced another factor of 30.  We 
are investigating the possibility of
interchanging electronically the top/bottom detector readout randomly
 about 100 times (in the course of the experiment) which brings us to
$3 \times 10^{-8}$ linearity error  
in the electronics.

 Alternatively we
will calibrate the system in situ by measuring the 
muon spin precession (without applying the radial E-field). The
resulting waveform should be an ideal 
sine-wave multiplied by an exponential decay.  Any discrepancy from this 
ideal waveform can be assigned to gain changes and corrected for.

\item{}  Proton or pion contamination of the muon beam.  During the 1997 run 
of the
g-2 experiment about $10^8$ pions per spill were injected, the 
pion flash was the biggest problem.   After striking a nucleus hadrons produce 
secondary neutrons which in turn produce gammas paralyzing the 
photo-multipliers mostly at the first half of the ring for tens of 
${\rm \mu s}$.  When we moved to 
muon injection by lowering the accepted
particle momentum by about $1.5\%$ the light flash went down by a factor
of 50.

In the muon EDM experiment the secondary beam is 900 MeV/c whereas the
muons have about 520 MeV/c.  One can expect a proton and pion leakage
of less than 
 $10^{-4}$ per muon.  It is therefore the lost muons (assuming $50\%$
storage efficiency) which will
dominate the flash.  Their efficiency ($\mu^+$ case) of producing
neutrons is down by approximately a factor of $\alpha \approx 1/137$ relative to
those produced by pions/protons.  We plan to study this effect with
the kicker off and observe its time dependence and up/down asymmetry.
  This effect is still under study.

\item{}  Muon losses as a function of time.  If there are muon losses
at an appreciable level and they are lost in a non-uniform way, they could
provide a false signal.  We are studying the potential level of this
effect and ways to minimize it (e.g. scraping, under-filling the
vertical and horizontal acceptance of the ring, etc.).

\item{}  Stored positrons.  Positrons from the beamline with the same
momentum as muons will be stored in the ring and will synchrotron
radiate their energy at a rate of 0.8~KeV per revolution.  For
$0.5\%$ momentum acceptance of the ring, it will take approximately
1~ms before all positrons reach the inner aperture whereas almost all 
the muons decay within $50\, \rm \mu s$.  We are studying
the potential problems and possible uses of the stored positrons.

\item{}  Due to the cylindrical nature of the electrostatic plates which 
provide the radial electric field, the E-field decreases as $1/R$ making
the cancelation of the g-2 imperfect as a function of the muon momentum.
Also, taking into account that the spin precession cancelation depends on 
both the magnetic field and the value of gamma 
(see equation~(\ref{eq:efield})), one can estimate that at the 
edge of the aperture, the muon spins will flip 
horizontally in $100 \, {\rm \mu s}$.   Only a very small fraction of the
muons will do that.
We are investigating the requirement on the acceptance and detector efficiency
uniformity so that this effect is also negligible.  At any rate this
error will be known by letting the muon spin precess and observing the
precession plane as a function of the E-field.

\item{} The problem with radial B-fields when using electrostatic 
focusing is eliminated completely because we use  only
magnetic focusing.  However the misalignment and time stability of the 
E and (with much less concern) B-fields will be monitored by the
inclinometer and the 
laser interferometers described in the next section.

\item{} Horizontal and Vertical coupling of the betatron
oscillations.  Yuri Orlov is studying this effect.

\item{} The ``vertical'' component of the E-field, i.e. the out of
plane component 
of the E-field should be known to $\pm 10^{-8}$ of E
for $1\times 10^{-24} \, \rm e \cdot cm$ limit.
This is addressed below.

\end{enumerate}

We feel confident that we can analyze all sources of systematic
errors and greatly reduce them due to the flexibility provided by a 
dedicated muon EDM experiment.

\subsection{Electrode alignment}
The vertical component of the electrostatic field in the muon storage region,
averaged over the ring circumference, must be well suppressed.  An average 
component of 10~nrad will induce an effect mimicking an EDM of 
$10^{-24}\,{\rm e \cdot cm}$.
This implies
that the electrodes, and the resistive top and bottom plates used to produce a
uniform electrostatic field, must be aligned with respect to a 
plane with high precision.

\begin{figure}[p]


\psfig{file=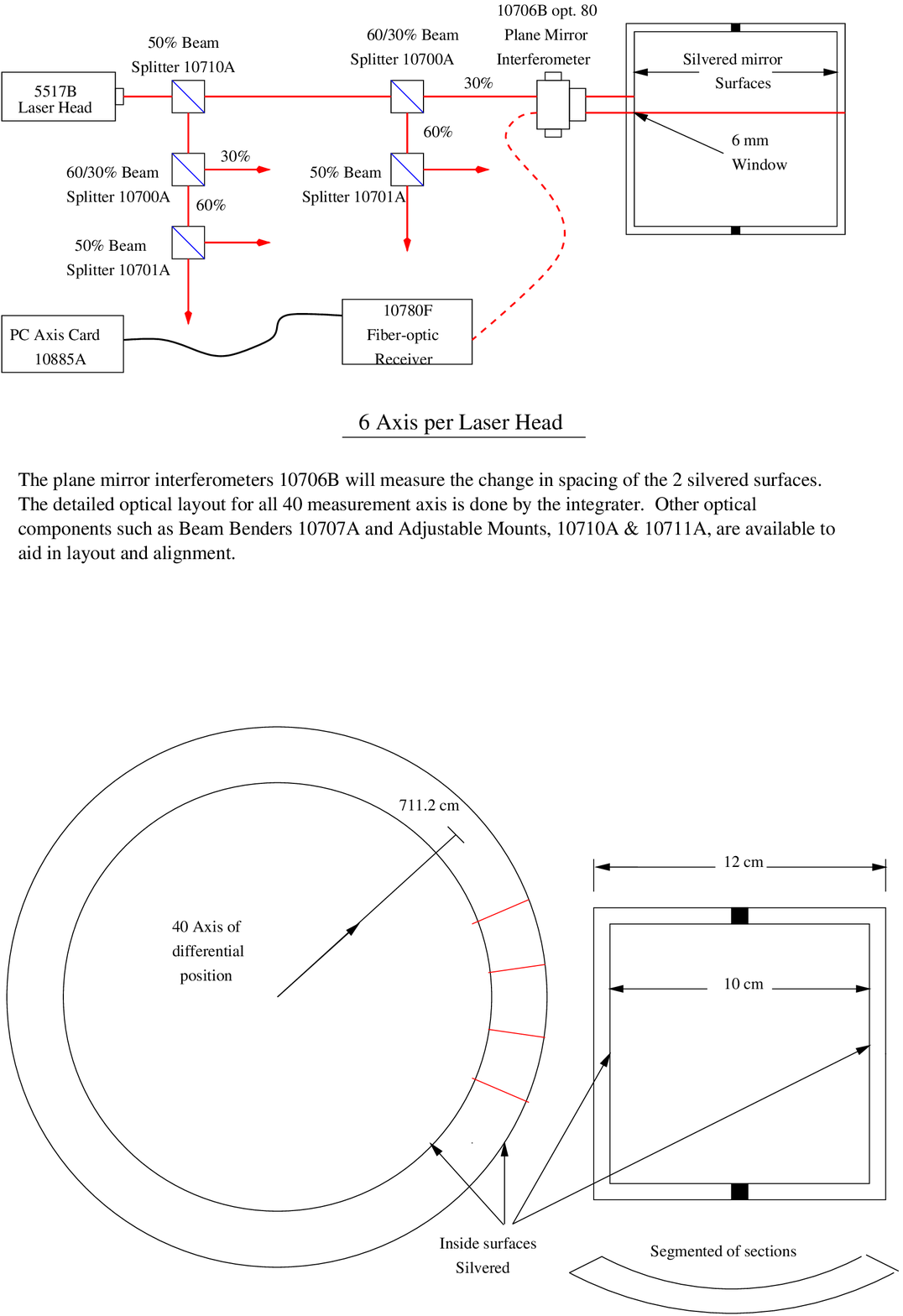,width=1.\textwidth}
\caption[hplaser]{{ The cross section of the plates (10~cm by
10~cm) which will also serve as the decay positron detectors, is shown
on the right.  The top and bottom plates consist of 
conductive glass plates.  On the top of the figure the laser
interferometer is shown which has 5~nm relative position resolution with 33~KHz
response speed.}
\label{fg:hplaser}}

\end{figure}

Fig.~\ref{fg:hplaser} shows a method that may be used to achieve this 
alignment and, most importantly, the monitoring of the relative distance 
to better than 10~nrad.  The
electrodes and top and bottom plates are clamped at intervals with pairs of
U-shaped machinable ceramic yokes.  The top and bottom plates are made of
glass doped to produce a small but uniform conductivity bonded to insulating 
webbing for support. The outer
surfaces of the conductive glass are covered with an anti-reflection coating so
that the parallelism of the inner surfaces of the top and bottom plates may be
tested, after clamping, by optical interferometry.  Shimming may be necessary
to insure that these inner surfaces are parallel.  We are
investigating the possibility of constructing the plates of scintillating 
lead glass to be an almost full acceptance detector for the
decay positrons.
 The cross section of the plates (10~cm by
10~cm) is shown on the right.  The top and bottom plates consist of
conductive glass plates.  On the top of the figure the laser
interferometer is shown which has 5~nm relative position resolution and 33~KHz
response speed.  Integrating over 3~ms will result to 5~nrad angular 
resolution~\cite{hp}.

The Inclinometer technique with Deuteron Polarimeter is described in
the Spring  
2000 version of the Letter of Intent~\cite{muon_loi_97}.  It turns out 
that Deuterons are much more sensitive to the out of plane component of
the E-field than muons are. The amazing coincidence is that although
the muon and deuteron anomalous magnetic moment (a) and $\gamma$ differ 
by orders of magnitude, the B-field required for storage is the same
to 0.015T for $\mid E 
\mid =$~2MV/m  and R = 7.11m (the muon g-2 ring radius).

\section{Summary}
We are proposing to search for an EDM of the muon by performing a dedicated
experiment in the existing g-2 ring.  The main differences from the g-2
experiment  are that we are
going to place the detector calorimeters inside  the vacuum 
chambers, use magnetic focusing (start with weak and continue with
strong later on) with no vertical electric fields 
present, lower momentum muons, and a lithium lens target with modified beam
transport and inflector system.  The introduction of an inclinometer,
i.e. a system of 
measuring in situ the out of plane component of the average radial
electric field direction (which is 
used to cancel the 
spin motion of the muon).    This is done  
by injecting polarized deuterons in the ring and observe their vertical
spin precession as a function of time.
Finally, the introduction of a monitoring system of the above
alignment of the electric and magnetic
field directions by a set of laser interferometers in between the
deuteron runs.

The estimated sensitivity is at the level of
$ 10^{-24} \, {\rm e \cdot cm} $,
an improvement of a factor of almost $10^5$ to $10^6$ over the last
CERN muon g-2  
experiment, with comparable improvement on the systematic errors.
This technique can may use of much more intense polarized muon sources
which will make it much easier to study the systematics and push them
further down to their limits.




\newpage

\end{document}